# High Mobility Ambipolar MoS$_2$ Field-Effect Transistors: Substrate and Dielectric Effects

Wenzhong Bao, Xinghan Cai, Dohun Kim, Karthik Sridhara, and Michael S. Fuhrer

Center for Nanophysics and Advanced Materials, University of Maryland, College Park, MD 20742-4111, USA

We fabricate MoS$_2$ field effect transistors on both SiO$_2$ and polymethyl methacrylate (PMMA) dielectrics and measure charge carrier mobility in a four-probe configuration. For multilayer MoS$_2$ on SiO$_2$, the mobility is 30-60 cm$^2$/Vs, relatively independent of thickness (15-90 nm), and most devices exhibit unipolar n-type behavior. In contrast, multilayer MoS$_2$ on PMMA shows mobility increasing with thickness, up to 470 cm$^2$/Vs (electrons) and 480 cm$^2$/Vs (holes) at thickness ~50 nm. The dependence of the mobility on thickness points to a long-range dielectric effect of the bulk MoS$_2$ in increasing mobility.

High quality two-dimensional materials have attracted significant attention due to their interesting physics and potential applications for electronic devices.[1] An outstanding example is graphene which has taken both the scientific and technological communities by storm[1-3]. However, the absence of a bandgap[4,5] inhibits its broader applications such as CMOS-like logic devices. Unlike graphene, $MoS_2$ has a thickness-dependent bandgap of 1.3-1.9eV[6-10] and is therefore promising for field-effect transistors (FETs). Indeed, recent studies of exfoliated thin $MoS_2$ on $SiO_2$ have demonstrated FE operation with high on-off ratios at room temperature.[11-16] The 1.9eV direct band gap of monolayer $MoS_2$ makes optoelectronic devices possible.[8,17,18] On the other hand, bilayer and multilayer $MoS_2$ devices are also interesting: the bandgap of bi-layer $MoS_2$ is expected to be tunable by vertical electric field,[19,20] and multilayer $MoS_2$ is expected to carry higher drive current than monolayer $MoS_2$ due to its lower band gap and triple of density of states at the conduction band minimum.[15,16] A recent report of gate-tuned superconductivity at temperatures up to 9.4 K in multilayer $MoS_2$ adds further interest.[21]

The substrate and overlayer of a thin-film transistor may affect its performance through the introduction of disorder (either long-ranged charge disorder, or short-ranged disorder caused by chemical bonding or roughness) which reduce the charge carrier mobility, and by dielectric screening which may enhance the mobility.[22] Previous studies of backgated exfoliated $MoS_2$ FET devices on bare $SiO_2$ found charge carrier mobility <50 $cm^2$/Vs and large subthreshold swings (>1 V/decade).[1,11,23] $MoS_2$ on $Al_2O_3$ substrates[15,16] showed improved mobility (>100 $cm^2$/Vs)[15] and good subthreshold swings (70-80 mV/decade).[15] Studies of $MoS_2$ on $SiO_2$ with various top gate schemes ($HfO_2$, $Al_2O_3$ and polymer electrolyte)[12-14] reported much higher mobility values (>900 $cm^2$/Vs) but it is likely that mobility was significantly overestimated in these dual-gated geometries.[24] However, it is also apparent that the addition of the superstrate dielectric enhanced the conductance of the $MoS_2$ devices. Hence, open questions remain as to the role of the dielectric substrate and overlayers in both causing and screening disorder in $MoS_2$ thin-film FETs.

In this paper, we report fabrication of MoS$_2$ FETs of varying thickness (1-80 nm) on both SiO$_2$ and on polymethyl methacrylate (PMMA) dielectric substrates. For multilayer MoS$_2$ on SiO$_2$, the mobility is on order of 30-60 cm$^2$/Vs consistent with previous results,[11] relatively independent of thickness, and most devices exhibit unipolar *n*-type behavior. In contrast, multilayer MoS$_2$ on PMMA shows mobility increasing with thickness, up to 470 cm$^2$/Vs (electrons) and 480 cm$^2$/Vs (holes) at thickness ~50 nm. The dependence of the mobility on thickness for thicknesses up to 80 nm is unexpected, and points to a long-range dielectric effect of the bulk MoS$_2$ in increasing mobility. Addition of a PMMA layer on top of the MoS$_2$ devices further increases the mobility, confirming the dielectric effect.

The starting substrates for device fabrication are 300 nm SiO$_2$ on Si, or spin-coated 300 nm PMMA on 300 nm SiO$_2$ on Si baked at 170°C for 30 min. We choose PMMA as a polymer dielectric because of its easy deposition by spin coating, low values for the trapped charge density and high dielectric constant similar to that of silicon dioxide (3.9 at 60 Hz). Thin films of MoS$_2$ were obtained by tape-cleavage of a single crystal geologic specimen of molybdenite (verified by both energy-dispersive X-ray spectroscopy and X-ray photoelectron spectroscopy to be single crystal 2H-type MoS$_2$) followed by mechanically exfoliation onto the device substrate.[11] Thin MoS$_2$ flakes are optically visible on such substrates, and among those we select the crystals that are as uniform as possible by examining optical microscope and atomic force microscope (AFM) images. Thickness *t* of the flakes is measured by AFM. Four-probe electrical contacts (30 nm Ti and 80 nm Al) are patterned on top of selected MoS$_2$ flakes using a shadow mask technique,[25] avoiding contamination by residues from chemical resists or developers. The conductivity $\sigma$ is measured in a four-probe configuration as a function of carrier density $n = c_g V_g$ where $c_g$ is the back-gate capacitance per unit area, and $V_g$ is the back-gate voltage. The potential difference across the voltage probes is kept below 100 mV in all measurements. In the four-probe configuration with a single back-gate electrode the charge-carrier mobility is well approximated by the field-effect mobility $\mu_{FE} = \frac{1}{e}\frac{d\sigma}{dn}$ where *e* is the elementary charge. By examining both two-probe and four-probe conductivity measurements (electrical setup as shown in Fig 1a), we observe that the contact resistance varies from sample to sample, and is especially large (~MΩ) for thin flakes (< 3 nm); the four-probe geometry eliminates contributions from contact resistance which could cause the conductance to be significantly underestimated.

Fig. 1a shows a schematic of our four-probe devices, and Fig. 1b shows an optical micrograph of a typical PMMA-supported $MoS_2$ device. Figures 1c and d display the room temperature four-probe conductivity $\sigma$ as a function of applied back gate voltage $V_g$ for four devices with thickness of 1.5, 6.5, 47 and 80-nm respectively, which summarize the range of observed device behaviors and demonstrate the qualitative trends observed with increasing device thickness. The 1.5-nm-thick device (bilayer $MoS_2$) displays an *n*-type unipolar behavior indicated by turning on of conductivity at positive $V_g$ (accumulation of electrons), while staying off at a window of negative $V_g$, which agrees with previously reported $MoS_2$ FETs on $SiO_2$. The 6.5 nm thick device shows clear ambipolar behavior in Fig 1d, but with much lower mobility (~1 $cm^2$/Vs) for holes compared to electrons (68 $cm^2$/Vs). Note that in the four-probe setup, the lower mobility is not an artifact due to contact resistance, but reflects the bulk carrier mobility. The 47-nm-thick device shows good ambipolar behavior with high mobility for electrons and holes (~270 and 480 $cm^2$/Vs, respectively). A trend toward a smaller window of off state (separating the electron and hole conduction regions) and higher off-state conductance is observed with increasing thickness, and by $t = 80$ nm there is no clear electron conduction, off, and hole conduction regions indicating that carriers in the bulk of the $MoS_2$ are likely dominating the conductance.

Figure 2 plots room temperature $\mu_{FE}$ as a function of thickness *t* for all measured $MoS_2$ flakes. To shed light on the ambipolar behavior and thickness dependence of $MoS_2$ devices, we systematically investigated more than 50 PMMA-supported devices and 6 $SiO_2$-supported devices, with thickness spanning from monolayer to 150 nm. We exclude the thick $MoS_2$ devices that show bulk behavior (e.g. $t = 80$ nm in Fig. 1c and d) and include only devices with a clear off state. Devices with ambipolar performance are indicated as dashed-line connected hollow squares (corresponding hole-carrier mobility) and solid squares (corresponding electron-carrier mobility). To avoid dielectric breakdown, the range of $V_g$ for PMMA-supported devices is ±150V and ±75V for SiO$_2$-supported devices. Most of the PMMA-supported thin devices (monolayer – 30 nm) and all SiO$_2$-supported devices display *n*-type unipolar behavior, while ambipolar performance is observed in most PMMA-supported thick devices (40 nm < *t* < 70 nm) with high-mobility,

The mobility of PMMA-supported MoS$_2$ devices show an increasing trend with thickness, up to 470 cm$^2$/Vs (electrons) and 480 cm$^2$/Vs (holes) for thicknesses near 50 nm. In most devices with ambipolar behavior, the hole-carrier mobility is larger than that of electrons, suggesting that *p*-type operation is more favorable for multilayer MoS$_2$ FETs. In contrast, the mobility of SiO$_2$-supported devices is much lower (30-60 cm$^2$/Vs) and almost thickness independent, and ambipolar behavior is rarely observed in them. Our observed mobilities are the highest four-probe room-temperature mobilities measured in MoS$_2$, comparable to the room temperature intrinsic hole mobility in silicon, and somewhat higher than existing estimates of phonon-limited room temperature mobility for thick[15] and single-layer MoS$_2$.[26] Notably, the highest electron and hole mobilities are comparable, an advantageous property for CMOS, possibly reflecting the similar electron and hole masses in MoS$_2$.[9] The observed mobilities are also the highest we are aware of in any ambipolar thin-film field-effect device with the exception of graphene.

The observation of a thickness-dependent mobility in MoS$_2$ up to thickness of 50 nm or more is surprising. Confinement effects on the bandgap are negligible beyond a few nm thickness,[8-10] and the carriers in MoS$_2$ FETs are expected to be confined with a few nm of the gate dielectric interface.[15] The thickness dependence points to a role for the additional MoS$_2$ above this region which contains few charge carriers. One possible explanation is that the additional MoS$_2$ layers serve as a dielectric capping layer which enhances screening of long-range disorder. To test this hypothesis, we add an additional dielectric layer (spin-coated PMMA, thickness of 300 nm) to most of the measured devices shown in Fig 2 and re-measured the mobility. The results are plotted in Fig 3a (only the higher of hole or electron mobility is plotted for devices with ambipolar behavior). Top-coating with PMMA enhances the mobility of most PMMA-supported devices, even those of thickness ~60 nm (maximum improvement of more than 300%), while $SiO_2$-supported devices only show a slight increase of mobility. Typical device output characteristics before and after double layer PMMA coverage are shown in Fig 3b and c for PMMA-supported and $SiO_2$ supported $MoS_2$, respectively. The results confirm that additional dielectric layers can enhance the mobility of $MoS_2$ on PMMA even for $MoS_2$ thicknesses of 60 nm, demonstrating the importance of long-range disorder in determining the mobility of $MoS_2$ on PMMA. For lower-mobility $MoS_2$ on $SiO_2$, the effect of increasing $MoS_2$ thickness or adding additional PMMA dielectric is small, indicating the dominance of short range disorder, likely chemical bonding to $SiO_2$ or interface roughness scattering.

In conclusion, we study the dependence of $MoS_2$ field effect mobility on substrate ($SiO_2$ and PMMA), $MoS_2$ thickness, and PMMA dielectric overlayer. $MoS_2$ on $SiO_2$ shows typically unipolar $n$-type behavior, low mobility relatively independent of $MoS_2$ thickness or dielectric overlayer. PMMA-supported devices show ambipolar behavior with the highest measured room temperature four-probe mobilities for $MoS_2$, increasing with thickness (comparable to $p$-type silicon; up to 470 $cm^2$/Vs for electrons and 480 $cm^2$/Vs for holes at thickness ~50 nm). For $MoS_2$ on PMMA mobility could often be improved even further by covering device surface with an extra top layer of PMMA. The strong dielectric effects on mobility for $MoS_2$ devices on PMMA imply a dominance of long-range disorder, while the absence of such effects for $MoS_2$ on $SiO_2$ implies a dominance of short-range disorder at the $SiO_2$ interface due to chemical bonding or surface roughness.

**Acknowledgments**

This work was supported by NSF Grant No. DMR-11-05224 and the US Office of Naval Research MURI program.

**Figure Captions**

FIG. 1. (a) Schematics of four-probe $MoS_2$ devices on $SiO_2$/Si and PMMA/$SiO_2$/Si. (b) Optical image of a typical PMMA-supported $MoS_2$ device. Green area is $MoS_2$ flake, yellow areas are Ti/Al electrodes, and blue area is the PMMA/$SiO_2$/Si substrate. (c-d) Conductivity $\sigma$ as a function of gate voltage $V_g$ on linear (c) and semi-logarithmic (d) scales for four $MoS_2$ devices on PMMA with thickness of 1.5 (red), 6.5 (brown), 47 (green) and 80 nm (blue).

FIG. 2. Room temperature field effect mobility $\mu_{FE}$ as function of thickness $t$ for 25 PMMA-supported (blue squares) and 6 $SiO_2$-supported $MoS_2$ (red circles) devices. Only electron mobility is shown for $SiO_2$-supported devices. PMMA-supported devices with measurable ambipolar behavior are indicated as dashed-line connected hollow squares (corresponding hole-carrier mobility) and solid squares (corresponding electron-carrier mobility).

FIG. 3. (a) Thickness-dependent field effect mobility $\mu_{FE}(t)$ of PMMA supported $MoS_2$ before (hollow triangles) and after (solid triangles) additional PMMA top-coating for most devices measured in Fig. 2. Only the higher of hole or electron mobility is plotted for devices with ambipolar behavior. (b-c) Typical $\sigma(V_g)$ characteristics before (red curve) and after (blue curve) PMMA top-coating for two PMMA-supported devices (b) and two $SiO_2$-supported devices (c). Insets: schematic views of devices after top-coating of PMMA.

Fig 1

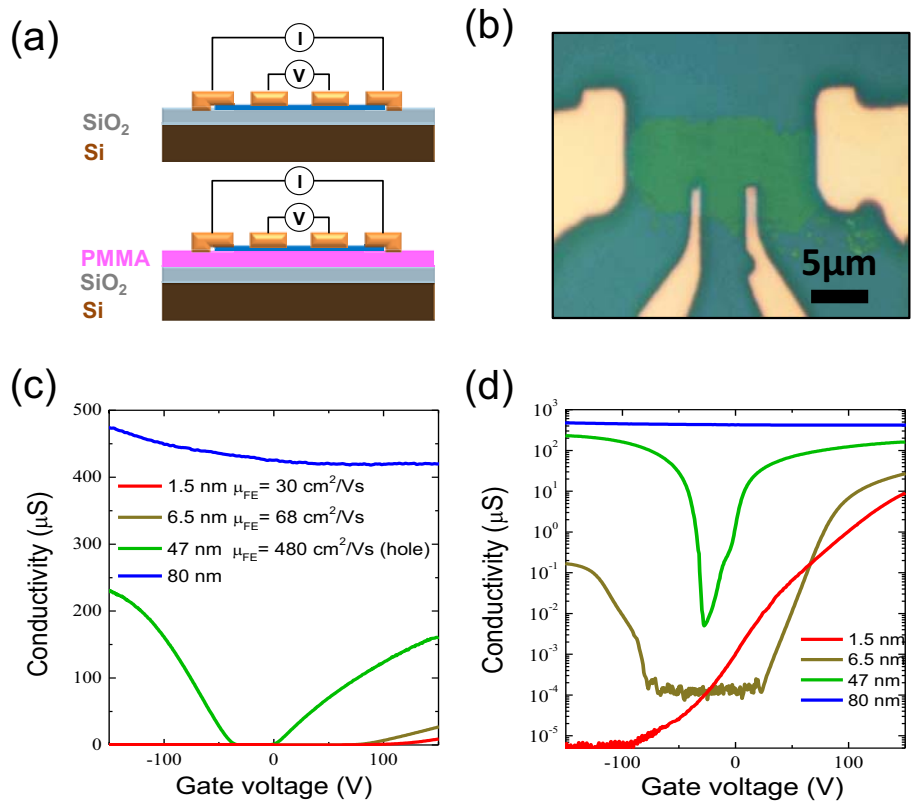

Fig 2

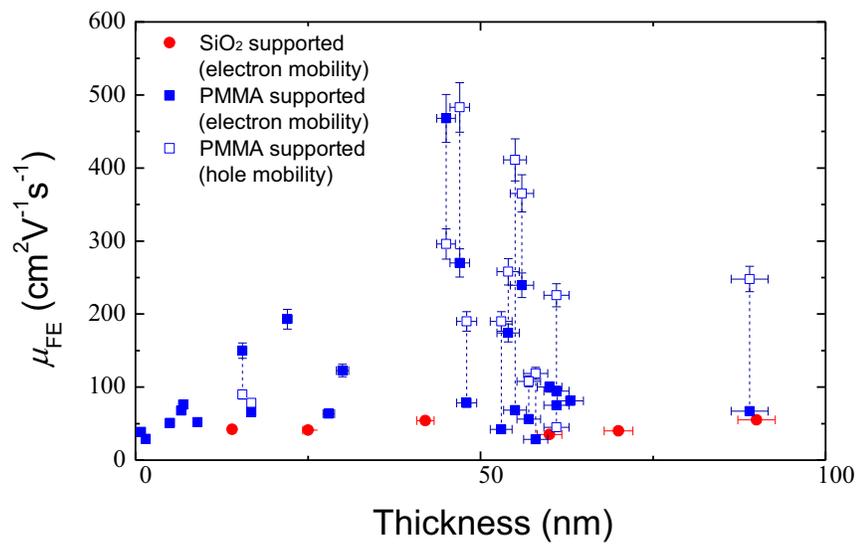

Fig 3

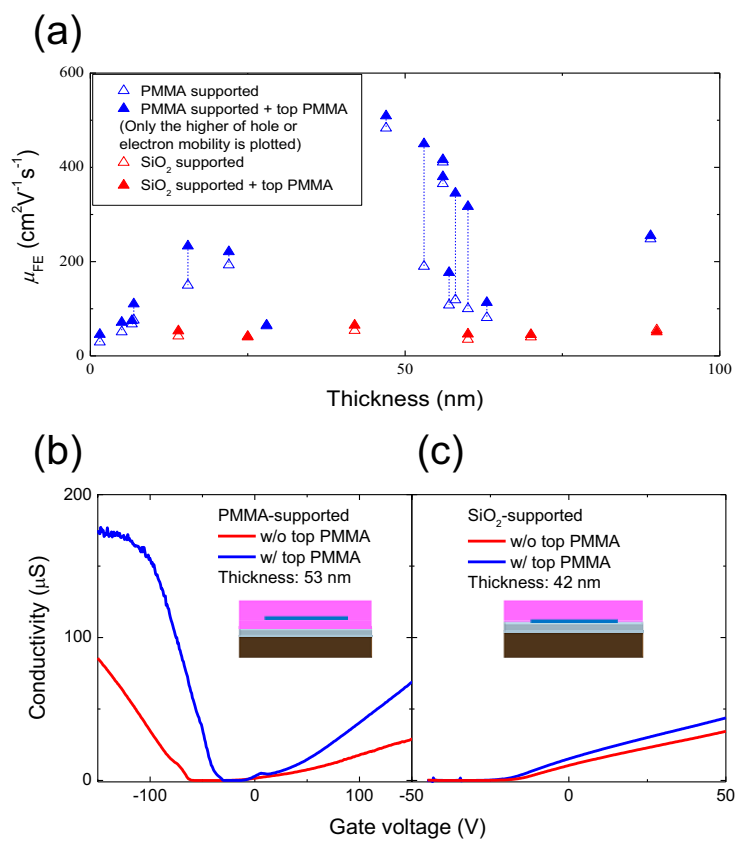